# Phase noise accumulation in recirculating frequency shifting loop based programmable optical frequency comb

ZHAOYU LU*, QUAN YUAN, AND DIANNAN HU

*Key Laboratory of the Ministry of Education on Optoelectronic Information Technology, School of Precision Instrument and Optoelectronics Engineering, Tianjin University, Tianjin 300072, China*
*luzhaoyu@tju.edu.cn

**Abstract:** The phenomenon of linewidth continuously broadening along with the recirculation number in recirculating frequency shifting loop is observed. In this paper, a novel method of measuring the phase noise accumulation induced by EDFA in recirculating frequency shifting loop is proposed and the experiment results support the viewpoint of the laser linewidth will be broadening by EDFA. An empirical formula has been extracted to estimate the linewidth deterioration of the RFS output. We demonstrate the relationship between the linewidth and the recirculation number of the RFS based optical frequency comb in both theoretically and experimentally. By employing the recirculation frequency shifting loop, the phase noise accumulation induced by EDFA could be measured obviously and the linewidth of each tone can be measured precisely.

## 1. Introduction

In the recent decade, the studies on programmable optical frequency comb (OFC) have received considerable attention. Programmable OFC in which the comb spacing and the center frequency are both tunable, as well as the adjusted number of comb-line, is widely used in carrier generation for multichannel coherent communications [1], remote sensing [2], spectroscopy and metrology [3]. There are several methods for programmable OFC generation, including cascaded phase modulators and intensity modulators [4], binary phase sampling modulation [5], nonlinear mixing in micro-resonators [6] and recirculating frequency shifting (RFS) [1,7]. Compared with these methods, the last method presents many advantages, such as continuously tunable comb spacing, lower radio-frequency (RF) driving voltage, and an intrinsic ability to achieve a large number of shifted carriers. Furthermore, it is a powerful technique for the generation of optical frequency comb and high repetition rate pulse with fundamental mode about 10 GHz [8]. At last, it is noticed that, unlike the mode-locking frequency combs, the programmable comb generation process is taking place outside of a resonant cavity, so that the individual comb teeth which are generated in this RFS scheme are not phase locked. Therefore, any tone in the comb can be easily frequency-locked to a standard RF oscillator, and it also makes adjusting the tone number and the comb spacing dynamically possible [9].

The performance of RFS based frequency comb is the key issue for its applications. The manifestation of the intensity noise accumulation in RFS loop based OFC has been intensively studied both theoretically and experimentally [10-12], since in some applications such as the multi-carrier communication, the signal to noise ratio (SNR) is the crucial criterion to evaluate the system performance. So mostly discussions had focused on the intensity noise and the intensity noise suppression [13] from amplified spontaneous emission (ASE) [14]. However, the phase noise characteristic is also a critical feature for modulation and demodulation of the signal with advanced modulation formats [15]. In FMCW interferometric experiment, the reduction in phase noise of the laser source leads to a better signal-to-noise ratio (SNR), by improving its coherence [16]. In Talbot laser, the frequency stability of each component plays a determining role on the generation of the high repetition rate pulse [17]. And in some other applications, such as remote sensing, spectroscopy and metrology, the stability of each tone's frequency is the most valuable indicator to evaluate the OFC quality. It is inferred that the linewidth of each tone is an ideal parameter to reveal the frequency stability of OFC.

It is generally recognized that the external modulation and the amplification process by erbium doped fiber amplifier (EDFA) will not induce additional phase noise [18] which means these operations will not cause the linewidth broadening of each tone after circulations. But in reality, the phase noise of each tone definitely accumulates and leads to a linewidth broadening somehow due to the ASE accumulation when each tone circling in the RFS loop [19]. The mechanism for the linewidth broadening of the output from an EDFA is generally considered as that the spontaneously emitted photons in the EDFA fall into the input laser linewidth after amplifying [20].

In this paper, the phenomenon of linewidth continuously broadening along with the increased recirculation number in the RFS loop is observed. A well-designed experiment scheme has been carrying out so that a mapping between the linewidth of each tone and its circulation number could be established by the experiment equipment. So we propose and demonstrate a relationship between the linewidth and the recirculation number in both theoretically and experimentally. Under the present model, we can predict that the phase noise of individual tone will accumulate along with the recirculation continuously and does not converge to a certain level.

## 2. Present schemes and theoretical analysis of phase noise characteristics

### 2.1 The programmable OFC generation and linewidth measurement schemes

The programmable OFC generator based on an RFS loop with a linewidth measurement scheme is illustrated in Fig. 1. A continuous wave (CW) laser is used as the seeding light source. A single sideband (SSB) modulator which is also called in-phase/quadrature (I/Q) modulator serves as a frequency shifter while the modulator is working on the single sideband modulation mode [21]. A bandpass filter (BPF) is used to control the number of tones recirculating in the RFS loop, and the frequency spacing $B$ of adjacent tones is equal to the frequency $f_m$ of the I/Q driven signals. The total loss in the loop is compensated by an erbium-doped fiber amplifier (EDFA) so that the frequency shifted light could grow up to an OFC after many times recirculation.

In order to quantitatively reveal the linewidth deteriorating of each tone influenced by ASE noise accumulation induced by EDFA, a tunable filter (TF) with narrow passband and a delayed self-heterodyne interferometer (DSHI) scheme [22] is adopted to measure the linewidth of each individual tone. The tunable filter we used has a rectangular passband profile with steep edges so that each individual tone of the RFS loop output can be filtered out for the following procedure of the linewidth measurement. Therefore the linewidth of each tone can be measured precisely without distribution by its adjacent tones.

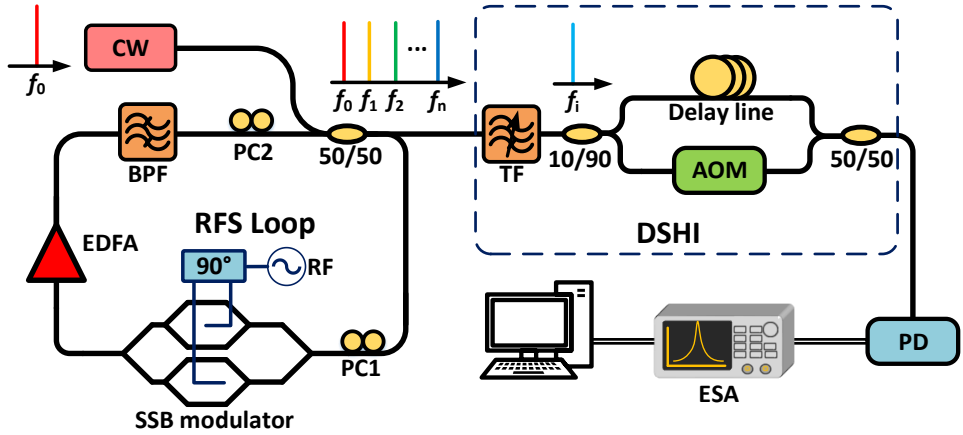

Fig. 1. Schematic of OFC generator based on the RFS loop and a DSHI linewidth measurement scheme. CW: continuous wave laser; PC: polarization controller; SSB: single sideband, RF: radio frequency generator; 90°: 90° hybrid coupler; EDFA: erbium doped fiber amplifier; BPF: band pass filter; TF: tunable filter; AOM: acoustic-optic modulator; PD: photodetector; ESA: electronic spectrum analyzer.

*2.2 Analysis of phase noise accumulation in the RFS loop*

Let's consider the phase noise induced by spontaneous emissions in a laser cavity firstly. The CW light wave generated from an actual laser source can be represented as

$$E_{in}(t) = E_0 e^{j[2\pi f_0 t + \varphi_0(t)]} \tag{1}$$

where $E_0$ is the amplitude of the electric field of the seeding light, $f_0$ is the nominal optical frequency and $\varphi_0(t)$ represents the random fluctuation of phase. The existing of phase fluctuation makes the laser not an ideal monochromatic field at frequency $f_0$, thus leads to a certain linewidth around $f_0$. The transfer function of the SSB modulator can be simply described as below by neglecting the high-order sidebands and its optical path [10]

$$T_{SSB}(t) = L_{SSB} e^{j[2\pi f_m t]} \tag{2}$$

where $L_{SSB}$ is the loss induced by the modulator operating in an SSB modulation mode, $f_m$ is the modulation frequency. The gain provided by the EDFA in the loop compensates equally to the losses induced by the SSB modulator and other insert components in the system. So the EDFA introduces additional ASE phase noise into the amplified signal relative to its input signal. This can be expressed as

$$G_{EDFA}(t) = G e^{j\varphi_{ASE}(t)} \tag{3}$$

where $G$ is the gain and $\varphi_{ASE}(t)$ is the changes in the phase induced by spontaneous emission in the EDFA.

Assuming the bandpass filter inside the RFS loop will prevent the light components which recirculating times over $n$, and the $L_R$ means the total loss in the RFS loop except for the SSB modulator. The total output $E_1(t)$ from RFS loop after the first round can be shown as

$$\begin{aligned} E_1(t) &= E_0(t) + T_{SSB}(t) G_{EDFA}(t) L_R E_0(t - \tau_R) \\ &= \frac{\sqrt{2}E_0}{2} e^{j[2\pi f_0 t + \varphi_0(t)]} + L_{SSB} L_R G \frac{\sqrt{2}E_0}{2} e^{j[2\pi (f_0 + f_m) t - 2\pi f_0 \tau_R + \varphi_0(t - \tau_R) + \varphi_{ASE}(t)]} \end{aligned} \tag{4}$$

in which $\tau_R$ means the time delay induced by the optical path of the RFS loop. So that the total output after $n$-th round from RFS loop $E_n(t)$ can be represented as

$$E_n(t) = E_0(t) + T_{SSB}(t)G_{EDFA}(t)L_R E_{n-1}(t-\tau_R)$$

$$= \frac{\sqrt{2}E_0}{2}e^{j[2\pi f_0 t + \varphi_0(t)]} + \ldots$$

$$+ (L_{SSB}L_R G)^n \frac{\sqrt{2}E_0}{2} e^{j[2\pi(f_0+nf_m)t - 2n\pi f_0 \tau_R - \pi n(n-1)f_m \tau_R + \varphi_0(t-n\tau_R) + \sum_{m=0}^{n-1}\varphi_{ASE}(t-m\tau_R)]}$$

$$= \frac{\sqrt{2}E_0}{2}e^{j[2\pi f_0 t + \varphi_0(t)]}$$

$$+ \sum_{k=1}^{n}(L_{SSB}L_R G)^k \frac{\sqrt{2}E_0}{2} e^{j[2\pi(f_0+kf_m)t - 2k\pi f_0 \tau_R - \pi k(k-1)f_m \tau_R + \varphi_0(t-k\tau_R) + \sum_{a=0}^{k-1}\varphi_{ASE}(t-a\tau_R)]}$$

(5)

In the above equation, the term including $f_0+kf_m$ corresponds to the $k$-th tone's frequency of output OFC, and the phase term of $\varphi_0(t-k\tau_R) + \sum_{a=0}^{k-1}\varphi_{ASE}(t-a\tau_R)$ corresponds to the accumulation of phase noise by amplified $k$ times in the EDFA. It is well known that the field corresponding to the spontaneous emission is not coherent with the amplified field by stimulated emission [23]. Therefore, the photons generated by spontaneous emission from an EDFA which continually join the amplified light field would interrupt the total laser field more frequently, the consequence is that the random phase fluctuations increase along with the spontaneous emission of the EDFA. So for the $k$-th tone of OFC output form RFS loop, its phase $\varphi_k(t)$ undergone $k$ times ASE noise is represented as

$$\varphi_k(t) = \varphi_0(t-k\tau_R) + \sum_{a=0}^{k-1}\varphi_{ASE}(t-a\tau_R) \tag{6}$$

The main reason of laser linewidth existent is that the phase fluctuation $\Delta\varphi(t,\tau)$ which can be described as $\varphi(t+\tau)-\varphi(t)$ is not constant. As the phase excursion $\Delta\varphi(t,\tau)$ is a net result of many small and statistically independent events, $\Delta\varphi_0(t,\tau)$ can be treated as Gaussian noise process obeying $\Delta\varphi_0(t,\tau) \sim (0,\sigma_0^2)$ when the central limit theorem of statistics applies. In a similar way, assuming $\Delta\varphi_{ASE}(t,\tau) \sim (0,\sigma_{ASE}^2)$ and the phase excursion $\Delta\varphi_k(t)$ can be expressed as

$$\Delta\varphi_k(t,\tau) = \varphi_k(t+\tau) - \varphi_k(t)$$

$$= \varphi_0(t-k\tau_R+\tau) - \varphi_0(t-k\tau_R) + \sum_{a=0}^{k-1}\varphi_{ASE}(t-a\tau_R+\tau) - \sum_{a=0}^{k-1}\varphi_{ASE}(t-a\tau_R) \tag{7}$$

$$= \Delta\varphi_0(t-k\tau_R,\tau) + \sum_{a=0}^{k-1}\Delta\varphi_{ASE}(t-a\tau_R,\tau)$$

According to the characteristic of Gaussian distribution, the variance of phase excursion $\Delta\varphi_k(t)$ is

$$\langle\Delta\varphi_k^2(t,\tau)\rangle = \langle\Delta\varphi_0^2(\tau)\rangle + k\langle\Delta\varphi_{ASE}^2(\tau)\rangle \tag{8}$$

It means the linewidth broadening induced by ASE noise is correlated linearly with the number of the recirculation times $k$ in the loop, since linewidth of the laser is directly correlated with its variance of phase excursion as Eq. (9).[23]

$$e^{-\pi\Delta v_k|\tau|} = e^{-\langle\varphi_k^2/2\rangle} \tag{9}$$

So we can suppose an empirical formula as Eq. (10). to describe the relationship between the linewidth $\Delta v_k$ of $k$-th tone with the shifted step of $B$ and the recirculation number $k$ to estimate the linewidth broadening after $k$ times recirculation in RFS loop.

$$\Delta v_k = \Delta v_0 + k \cdot \Delta v_{ASE} \tag{10}$$

where $\Delta v_0$ is the linewidth of seeding laser, and $\Delta v_{ASE}$ is the ASE noise-induced linewidth broadening in every circulation.

## 3. Experimental setup and results

The RFS loop based programmable OFC generator is illustrated in Fig. 2(a). A narrow linewidth of ~20kHz distributed feedback (DFB) laser is used as the seeding light source. The SSB modulator is driven by a 12.5GHz RF IQ signal so that a carrier suppression and frequency shifted +1 sideband signal can be observed after the SSB modulator, as shown in Fig. 2(b). A bandwidth of ~800 GHz bandpass filter controlled the seeding light recirculating 64 times in the RFS loop, so a 65 tones OFC with 12.5GHz comb spacing could be observed on the optical spectrum analyzer (OSA, Yokogawa AQ6370D), as shown in Fig. 3(a). The total loss in the RFS loop is compensated by an erbium-doped fiber amplifier (EDFA, IRE-POLUS EAD-60). With 10dB fluctuation of each tone's peak power, the generated OFC is not quite flat since the gain spectrum inequality of EDFA.

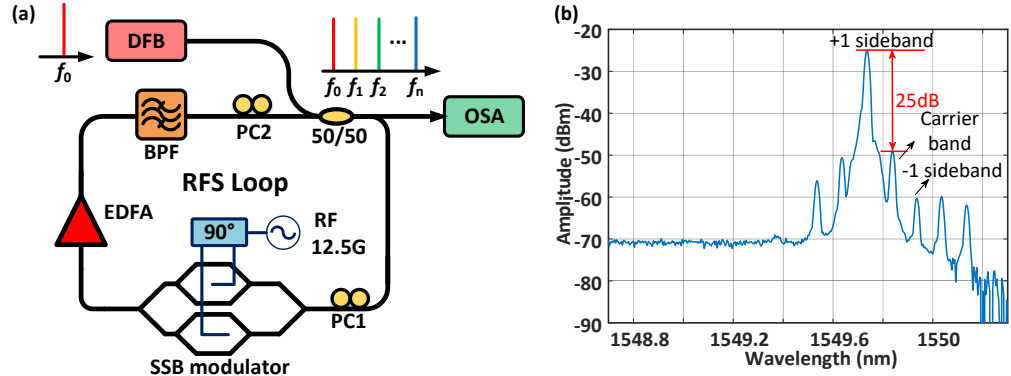

Fig. 2. Schematic of the RFS loop based OFC generator and a DSHI linewidth measurement scheme. CW: continuous wave light; PC: polarization controller; SSB: single sideband, RF: radio frequency generator; 90°: 90° hybrid coupler; EDFA: erbium doped fiber amplifier; BPF: band pass filter; TF: tunable filter; AOM: acoustic-optic modulator; PD: photodetector; ESA: electronic spectrum analyzer.

The linewidth of every single tone can be measured by DSHI method after filtering it out from the OFC spectrum by a tunable narrow bandwidth filter (Santec OTF-350). In order to achieve sufficient linewidth resolution, a length of 20km delay fiber was placed in the DSHI scheme. The measurement and Lorentz fitting results of the linewidth of the original incident laser (i.e. the $0^{th}$ tone), $30^{th}$ circulation output and the $60^{th}$ circulation output are shown in the Fig3. (d) (c) (b), respectively. It can be seen directly the linewidth of the OFC tone is increasing along with the circulation number in the RFS loop. But the deterioration of tone's linewidth is not a beneficial property of the OFC related applications. So it is necessary to figure out the relationship between linewidth and circulation number so that one can control the linewidth of OFC tones under an acceptable level.

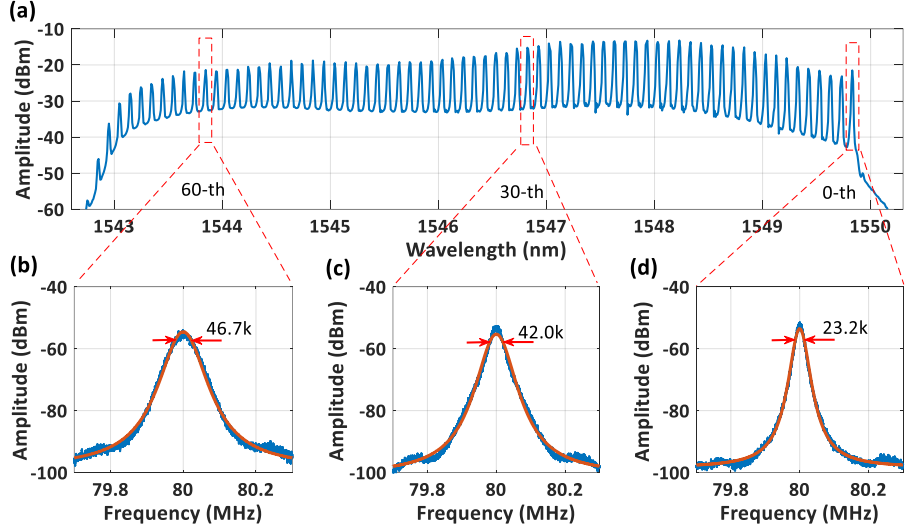

Fig. 3. (a) Measured spectrum of programmable OFC with comb spacing of 12.5G; (b) Measured linewidth about 46.7kHz of 60$^{th}$ circulation tone; (c) Measured linewidth about 42.0kHz of 30$^{th}$ circulation tone;(d) Measured linewidth about 23.2kHz of the original incident laser.

The same experiment under different frequency shifted step has been done which RF output frequency was set as 10GHz 15GHz and17.5GHz in the RFS loop respectively. And the linewidth of tone was recorded once every five circulations. The linewidths data under different shifted step was assembled in Table 1. It is clear that the linewidth of tone is increasing with the circulation number under the same shifted step.

Table 1. Measured linewidths under different shifted step with increasing loop number

| | Step(GHz) | | | | | Step(GHz) | | | |
|---|---|---|---|---|---|---|---|---|---|
| Loop($k$) | 10 | 12.5 | 15 | 17.5 | Loop($k$) | 10 | 12.5 | 15 | 17.5 |
| 0 | 23.2k | 23.2k | 23.2k | 23.2k | 45 | 38.7k | 43.6k | 48.9k | 49.4k |
| 5 | 27.0k | 28.8k | 26.6k | 27.3k | 50 | 39.2k | 44.5k | 45.3k | 51.6k |
| 10 | 28.5k | 33.6k | 28.7k | 32.0k | 55 | 41.6k | 45.0k | 49.2k | |
| 15 | 31.9k | 35.7k | 35.9k | 35.7k | 60 | 41.4k | 46.7k | | |
| 20 | 35.2k | 36.4k | 37.8k | 42.5k | 65 | 41.7k | | | |
| 25 | 35.3k | 40.5k | 40.5k | 43.0k | 70 | 44.9k | | | |
| 30 | 37.3k | 42.0k | 40.1k | 45.3k | 75 | 43.9k | | | |
| 35 | 36.8k | 41.8k | 42.2k | 47.7k | 80 | 43.8k | | | |
| 40 | 38.7k | 46.9k | 44.9k | 50.7k | | | | | |

As can be seen in Fig. 4, the linewidth of tone broadens almost linearly with the frequency shifting going on. The reason is that in the RFS loop, the tone whose frequency is farther away from the seeding light will circulate more times, and every time the light finished one loop, it suffered nearly same power gain from the EDFA and identical ASE noises. An incidental fact that we found is the linewidth broaden speed is related to the frequency shifting step which controlled by the RF source, and we consider that is caused by broader spontaneous photons mixed into each tone. Fitting results based on the empirical model mentioned above also overlay on Fig. 4. Under this model, it reveals the linewidth is proportional to the circulation number $k$ of this tone. Because the intensity of each tone is not equal, the measurement points of linewidth deviate to the linear fitting curve. Furthermore, a flattened spectrum programmable OFC will lead to a better linearity of tone's linewidth.

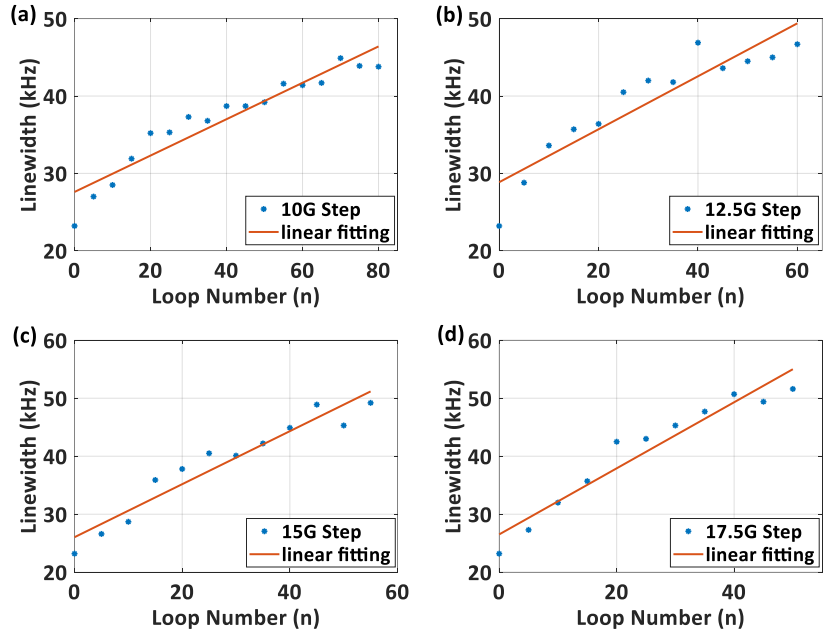

Fig. 4. Data point (blue dot) and correlated fitting curve (red line) of the relationship between the linewidth and loop number(*n*) under different frequency shifted step of (a)10GHz, (b)12.5GHz, (c)15GHz and (d)17.5GHz, respectively.

## 4. Conclusion

In this paper, a novel method of observing the phenomenon of phase noise accumulation induced by EDFA is proposed and new results are captured. By employing the recirculation frequency shifted loop, the phase noise accumulation induced by EDFA could be measured obviously and the linewidth of each tone can be measured precisely. A map between the frequency of each tone and loop number *k* is established by the RFS loop so that the phase noise induced by accumulated ASE noise can be obtained by measuring tones at a different frequency.

The phase noise accumulation induced by EDFA in the RFS loop and the linewidth broadens linearly with the recirculation number is demonstrated. An empirical formula has been extracted to estimate the linewidth deterioration of the RFS output. EDFA could cause the linewidth of amplified laser signal broadening is observed experimentally, this result will provide an effective tool to predict the limitation of cascade all-optical relay number in a long-haul optical communication system. What's more, the measurement result indicates that an approximate comb spacing should be set in the RFS loop to maintain a narrow linewidth characteristic of those tones far away from the seeding light.

## Funding

National Natural Science Foundation of China (NSFC) (61471256, 61575143, 61275084, 61377078) and Natural Science Foundation of Tianjin City (18JCYBJC16800).

## Acknowledgments

The authors would like to thank Prof. Ram Rajeev from Massachusetts Institute of Technology (MIT) for helpful discussions about the phase noise accumulation phenomenon.